\documentclass[final,5p,times,twocolumn]{elsarticle}

\usepackage{amssymb}

\usepackage{textgreek}
\journal{Nuclear Instruments and Methods A}

\begin{document}

\begin{frontmatter}

\title{Depleted Fully Monolithic Active CMOS Pixel Sensors (DMAPS) in High Resistivity 150~nm Technology for LHC}

\author[bonn,corr]{Toko Hirono} 
\cortext[corr]{Corresponding author}
\ead{hirono@physik.uni-bonn.de}
\author[cppm]{Marlon Barbero} 
\author[cppm]{Pierre Barrillon} 
\author[cppm]{Siddharth Bhat}
\author[cppm]{Patrick Breugnon} 
\author[bonn]{Ivan Caicedo} 
\author[cppm]{Zongde Chen}
\author[bonn]{Michael Daas} 
\author[irfu]{Yavuz Degerli}
\author[irfu]{Fabrice Guilloux}
\author[bonn]{Tomasz Hemperek} 
\author[bonn]{Fabian H\"ugging} 
\author[bonn]{Hans Kr\"uger} 
\author[cppm]{Patrick Pangaud} 
\author[bonn]{Piotr Rymaszewski} 
\author[irfu]{Philippe Schwemling} 
\author[irfu]{Maxence Vandenbroucke}
\author[bonn]{Tianyang Wang} 
\author[bonn]{Norbert Wermes} 

\address[bonn]{University of Bonn, Nussallee 12, 53121 Bonn, Germany}
\address[cppm]{Centre de physique des particules de Marseille, 163 Avenue de Luminy, Marseille, France}
\address[irfu]{IRFU, CEA-Saclay, Gif-sur-Yvette Cedex, 91191 France}

\begin{abstract}
Depleted monolithic CMOS active pixel sensors (DMAPS) have been developed in order to demonstrate their suitability as pixel detectors in the outer layers of a toroidal LHC apparatus inner tracker (ATLAS ITk) pixel detector in the high-luminosity large hadron collider (HL-LHC).
Two prototypes have been fabricated using 150 nm CMOS technology on high resistivity ($\geq$~2~k${\Omega}$$\cdot$cm$^2$) wafers.
The chip size is equivalent to that of the current ATLAS pixel detector readout chip.  
One of the prototypes is used for detailed characterization of the sensor and the analog readout of the DMAPS. 
The other is a fully monolithic DMAPS including fast readout digital logic that handles the required hit rate.
In order to yield a strong homogeneous electric field within the sensor volume, thinning of the wafer was tested.
The prototypes were irradiated with X-ray up to a total ionization dose (TID) of 50~Mrad and  with neutrons up to non-ionizing energy loss (NIEL) of 10$^{15}$~n$_{\rm eq}$/cm$^2$.
The analog readout circuitry maintained its performance after TID irradiation, and the hit-efficiency at $>$~10$^{-7}$ noise occupancy was as high as 98.9~\% after NIEL irradiation.

\end{abstract}


\end{frontmatter}


\section{Introduction}
\label{sec:intro}

Silicon pixel detectors in a toroidal LHC apparatus inner tracker (ATLAS ITk) in High-Luminosity Large Hadron Collider (HL-LHC) are expected to be used in high quantities \cite{atlas:letter}.
A monolithic CMOS active pixel sensor has sensing and readout functionalities integrated into one silicon piece. 
An advantage of monolithic CMOS active pixel sensors is the simplicity of their mass production process compared to conventional hybrid detectors, which require fine-pitch bump bonding between a sensor device and a readout chip.

In addition to the ease of volume production, the future experiments require high radiation hardness and high rate readout capability. 
The outer layers of the ATLAS ITk pixel detector are estimated to be exposed to total ionization dose (TID) of 50-80~Mrad and non-ionizing energy loss (NIEL) of 1-2$\times$10$^{15}$~n$_{\rm eq}$/cm$^2$.
Fast charge collection is mandatory to achieve these requirements \cite{tomasz:vertex}, i.e., depletion of sensor is required. 
Therefore, depleted monolithic CMOS active pixel sensors (DMAPS) have been developed for use in future ATLAS experiments.

\begin{figure}
    \centering   
    \includegraphics[width=0.8\linewidth,trim=0cm 12cm 4cm 0cm,clip]{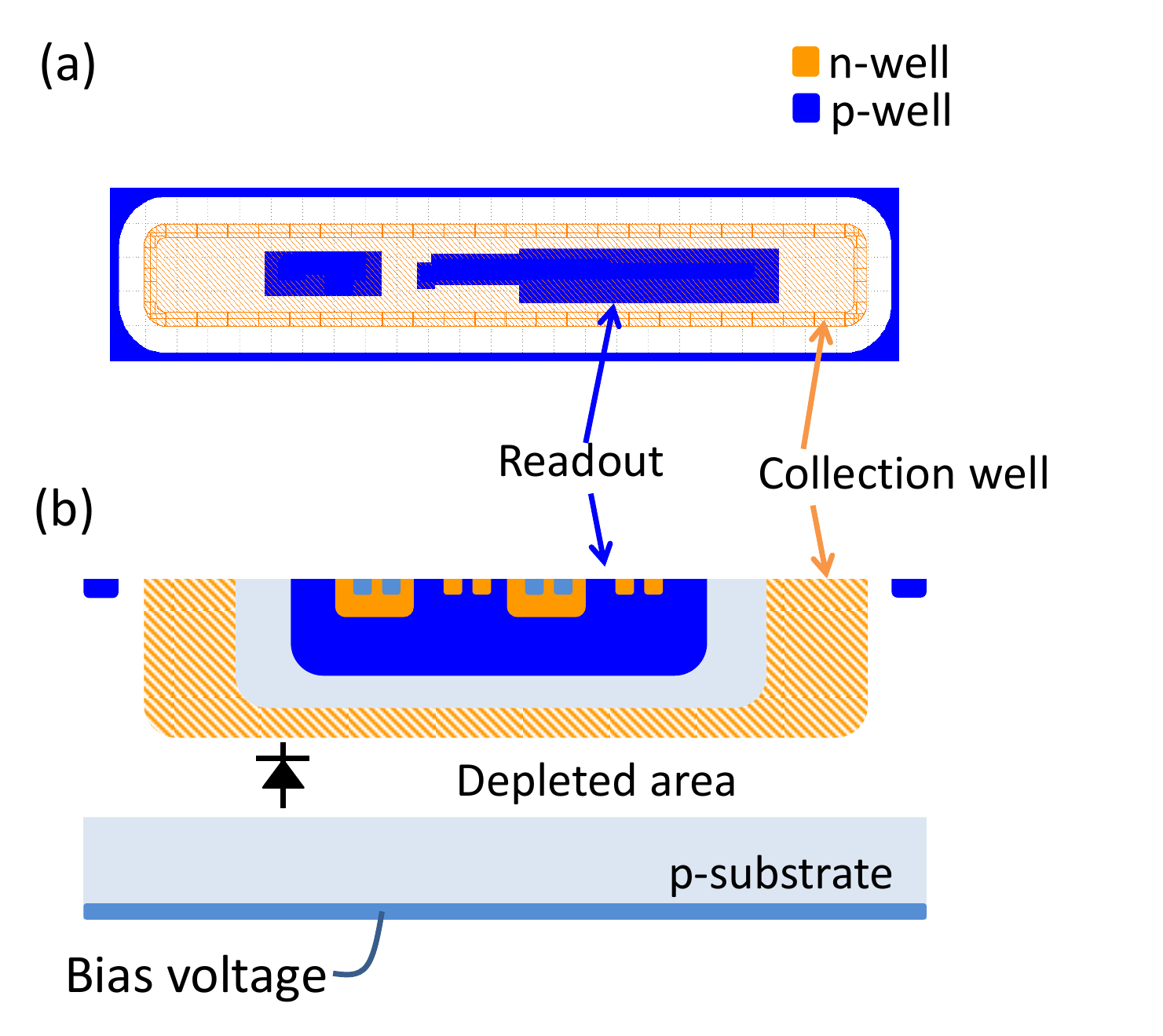}
    \caption[Top view and cross-section of the large fill factor design]{Top view (a) and cross-section (b) of the large collection node design. 
        The top view shows the actual layout of an LF-Monopix single pixel.
        The cross-section is a simplified drawing.
        The pn-junction created between the collection well and the substrate acts as a sensor, and is indicated by a sign of a diode.}
    \label{fig:layout}
\end{figure}

A DMAPS design that can apply a high bias voltage to the substrate was proposed \cite{ivan:2007, ivan:2015}.
Figure \ref{fig:layout} shows a top view and cross-section of a typical pixel in this design. 
Relying on a multi-well CMOS process, MOSFETs of the pixel readout are situated inside the charge collection node and are isolated from the substrate. 
Hence, it is possible to apply a high voltage (for instance, 200 V) to the substrate without damaging the readout.
The high bias voltage should deplete the sensor and yield a strong electric field, resulting in a high velocity of charge carriers even after NIEL irradiation. 

Another advantage of this design is the small gap between the charge collection nodes.
The design allows for small distances between the charge creation point and the collection node across the pixel.
This improves the radiation hardness.

DMPAS in a thinned high resistivity (HR) wafer are also expected to be radiation hardened.
Removing the unnecessary substrate increases the strength of electric field in the depletion region, and
the charge velocity depends on the resistivity of the depleted area. 
In addition, noise can be suppressed using HR wafer because of its small leakage current.


Based on the success of the first small HR large collection node DMAPS prototype \cite{hirono:2015, hirono:2016}, 
two prototypes have been developed to investigate the suitability of DMAPS as an ATLAS ITK pixel detector. 
This paper presents the radiation hardness of the HR large collection node DMAPS prototypes. 



\section{Prototypes}
\label{sec:proto}

Two prototypes, named LF-CPIX and LF-Monopix, were designed and fabricated using the LFoundry 150 nm CMOS process \cite{lf} on a high resistivity wafer ($\geq$~2~k$\Omega$$\cdot$cm$^2$).
The chip size and pixel size in both prototypes are 10~mm~$\times$~10~mm and 250~{\textmu}m~$\times$~50~{\textmu}m, respectively.
The chip size is larger than that of FE-I3, which is one of the current ATLAS hybrid pixel detector readout chips (FE-I4) \cite{fei3}. 
The pixel size is equal to that of the other current ATLAS hybrid pixel detector readout chip (FE-I4) \cite{fei4}.
The layout of a pixel of LF-Monopix is shown in Figure \ref{fig:layout} (a) 
The sensitive volume is the p-type substrate and the collection node is the n-well structure.
The size of the collection node is 230~{\textmu}m~$\times$~30~{\textmu}m, and hence the node gap between the pixels is 20~{\textmu}m. 

LF-CPIX has been developed for detailed studies of the sensor and analog readout.
The design of the guard rings of the LF-CPIX has been optimized \cite{Liu:2017} according to the results of the first small prototype \cite{hirono:2015}. 
Each pixel has a charge sensitive amplifier (CSA), a discriminator with 4 bits of trim DAC (TDAC), and a register (HIT register).
Three types of CSA are implemented in the LF-CPIX \cite{irfu:2016}. 
The CSA input device in each type is PMOS and NMOS (CMOS), PMOS, and NMOS transistors, respectively.
All CSAs are designed to have the same time walk and noise. 
The CMOS-CSA has the lowest power consumption among the three types. 
The PMOS-CSA has the largest power consumption, but it has been tested in many previous DMAPS prototypes and is confirmed to have acceptable radiation hardness \cite{ivan:2015, hirono:2016}.  
The NMOS-CSA has a simpler design than the CMOS-CSA and lower power consumption than the PMOS-CSA.
No fast readout is implemented on the LF-CPIX.
An image can be obtained using the HIT register, which stores hit/no-hit binary information coming from the in-pixel discriminator.

On the other hand, a pixel of LF-Monopix has logic for fast readout in addition to the circuitry in the LF-CPIX pixel \cite{tianyang:2017}.
In terms of hit rate, the environment of the inner layers of current ATLAS pixel detector is similar to that of the outer layers of the ATLAS ITk pixel detector. 
Therefore, the fast readout architecture used in Lf-Monopix is similar to that of FE-I3.
A pixel sends the address of the pixel in 14 bits and a time stamp of leading and trailing edge of the discriminator in 8 bits each.
The charge can be obtained in 8 bits resolution from the time over threshold (ToT) value, which is the difference between the time stamps of the trailing and that of leading edges.
The time stamp and readout clock frequencies were set to 40~MHz.


\section{Thinning and back-side process}
\label{sec:thin}
Substrate thinning in the LF-CPIX was performed from the back-side of the wafers down to 200~{\textmu}m or 100~{\textmu}m after CMOS fabrication.
The back-side was polished by etching, implanted dopant and annealed.
Then, metal contacts were deposited.

\begin{figure}
    \centering   
    \includegraphics[width=0.8\linewidth,trim=0cm 15cm 5cm 0cm,clip]{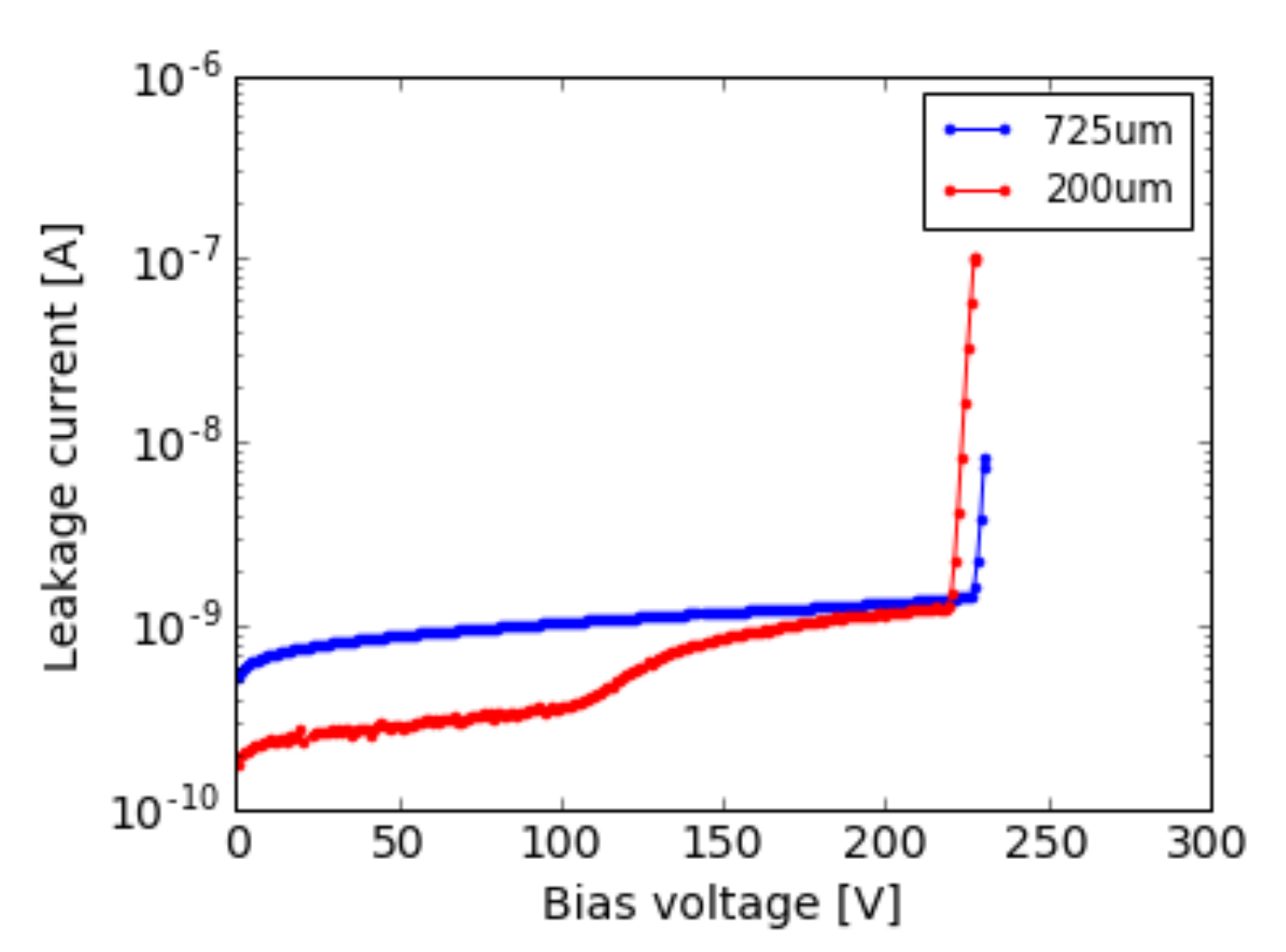}
    \caption[I-V curve of LF-CPIX with and without back-side process]{
        I-V curve of LF-CPIX with and withtout back-side processing.
        The blue curve shows the LF-CPIX leakage current without any thinning or back-side processing.
        The red curve shows the LF-CPIX leakage current after thinning down to to 200~{\textmu}m and performing back-side processing.}
    \label{fig:iv}
\end{figure}

A comparison between the leakage current of the 200~{\textmu}m thinned LF-CPIX and the un-thinned LF-CPIX is shown in Figure \ref{fig:iv}.
The breakdown voltage in both chips is higher than 220~V. 
There is no obvious lowering of the breakdown voltage due to the back-side process.
A small increase of leakage current is observed around the full depletion voltage ($\sim$ 120~V).
However, the increase is less than 1~nA and is negligible.

\begin{figure}
    \centering   
    \includegraphics[width=0.8\linewidth,trim=0cm 15cm 5cm 0cm,clip]{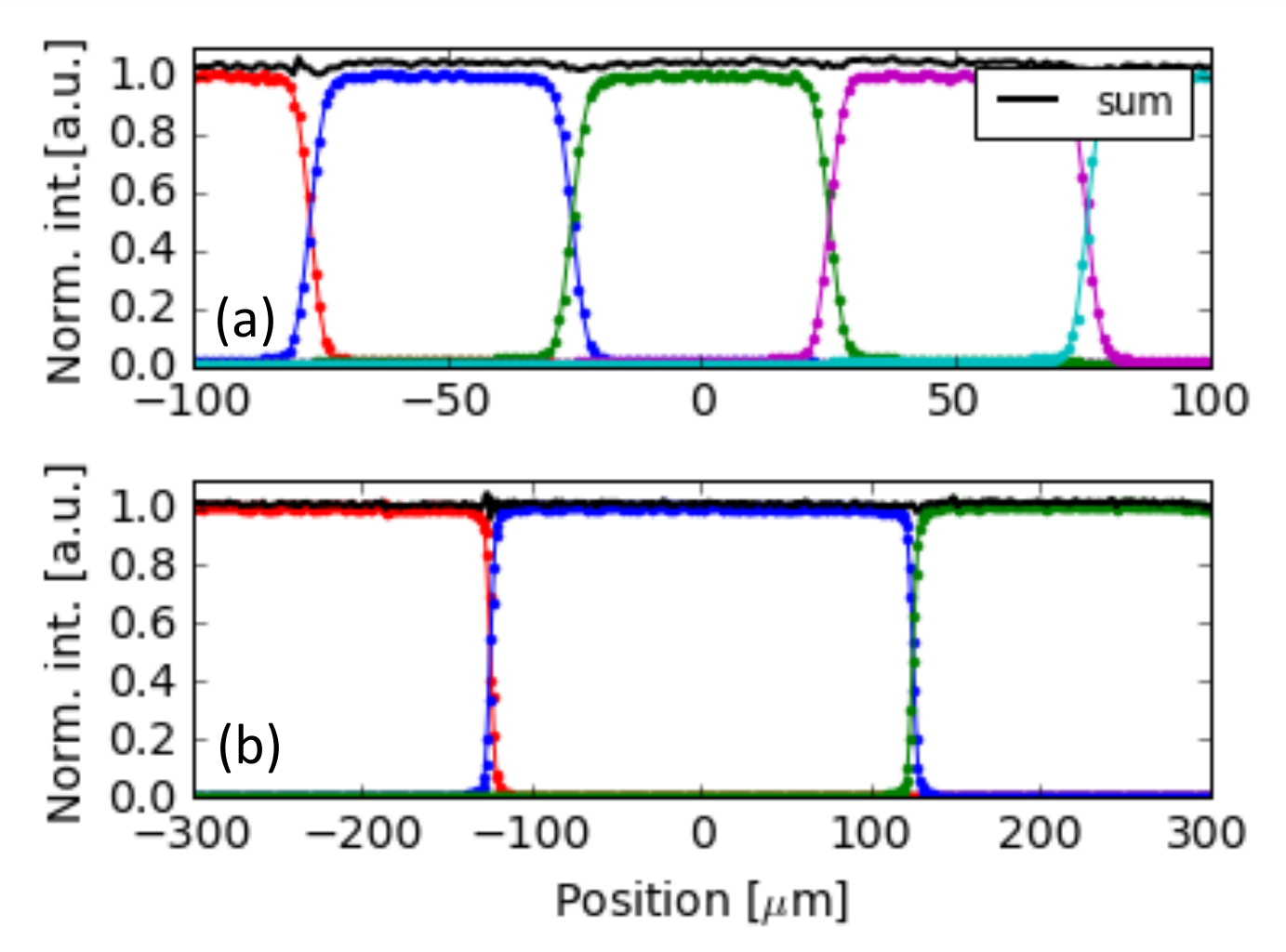}
    \caption[Laser response of LF-CPIX]{
    Laser response of LF-CPIX at the center of the long side across the short side (a) and that of the short side across the long side (b). Red, blue, green magenta, and cyan dots shows the the normalized measured intensity of the laser response from a pixel. The fitted lines are shown in the same color as the measured dots, accordingly. Black dots show the sum of the normalized laser response.}
    \label{fig:profile}
\end{figure}

Sensor homogeneity has been measured using a laser with 680~nm wavelength.
LF-CPIX was thinned down to 100~{\textmu}m in this test. 
Metallization was omitted in the back-side process in order to inject the laser.
Figure \ref{fig:profile} (a) and (b) shows the center of the long and short side of a pixel, receptively. 
The bias voltage is 200~V.
The profile of each pixel is fitted by a convolution of boxcar and Gaussian functions.
The measured data were normalized by the height of the fitted function.
The gain difference between each pixel is compensated by this normalization.
The sum of pixels is shown in Figure \ref{fig:profile}, and no in-pixel structures are observed from the laser response. 

\begin{figure}
    \centering   
    \includegraphics[width=0.8\linewidth,trim=0cm 15cm 5cm 0cm,clip]{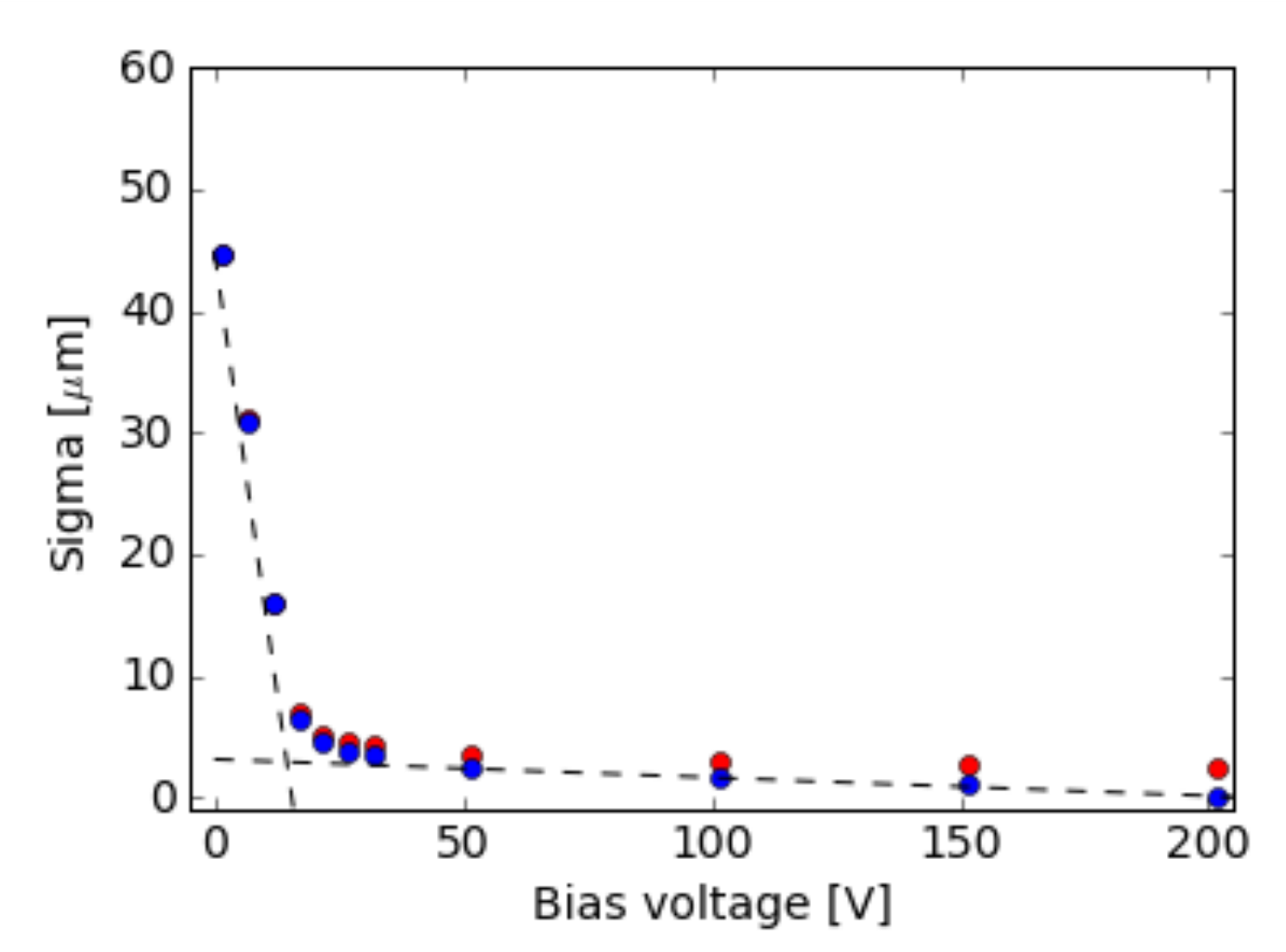}
    \caption[$\sigma$ of the pixel profile]{
    $\sigma$ of the pixel profile.
    The diameter of the laser spot is assumed to be 0~{\textmu}m (red dots) and 2.5~{\textmu}m (blue dots). 
    }
    \label{fig:bias}
\end{figure}

The variance of the Gaussian function ($\sigma$) is plotted in Figure \ref{fig:bias} against the bias voltage.
Charge diffusion during the transport from the back-side surface to the collection node is between the two values, which are calculated assuming the diameter of the laser spot is 0~{\textmu}m and 2.5~{\textmu}m, receptively. 
The measured $\sigma$ value is the square root of the sum of the squared laser spot size and the charge diffusion length.
In both cases, $\sigma$ saturates around a 15~V bias voltage and slowly decreases when the bias voltage is higher than 15~V. 
Since the attenuation length of the laser in silicon is 4~{\textmu}m, the result indicates that the full depletion voltage of the DMAPS is around 15~V. 
This is consistent with the resistivity as specified by the foundry ($\geq$~2~k$\Omega$$\cdot$cm$^2$).

\section{Radiation hardness}
\label{sec:res}

%
TID and NIEL irradiation tests were performed separately since each shows a different effect on the DMAPS. 

\subsection{TID radiation hardness}
Charge build-up occurs near the DMAPS surface due to TID \cite{pdg:2016}.
The analog readout, namely the CSA and discriminator, rely on transistor characteristics, such as gate threshold voltage or drain-source current.
Those characteristics are sensitive to charge build-up. 
LF-CPIX was irradiated to test the TID radiation hardness of the analog readout.
The Institute of Experimental Nuclear Physics Irradiation Center at the Karlsruhe Institute for Technology provided the X-ray tube \cite{kit} 
Irradiation was performed at a rate of 0.57 Mrad/h. 
The irradiation was paused 21 times so that measurements could be taken, where each measurement takes 1~hour.
The average temperature of the chip during irradiation was 27~$^\circ$C. 
Since the time dependency of the annealing effect during the measurement is not negligible, the same measurement procedure was repeated at each pause.

\begin{figure}
    \centering   
    \includegraphics[width=0.8\linewidth,trim=0cm 10.5cm 5cm 0cm,clip]{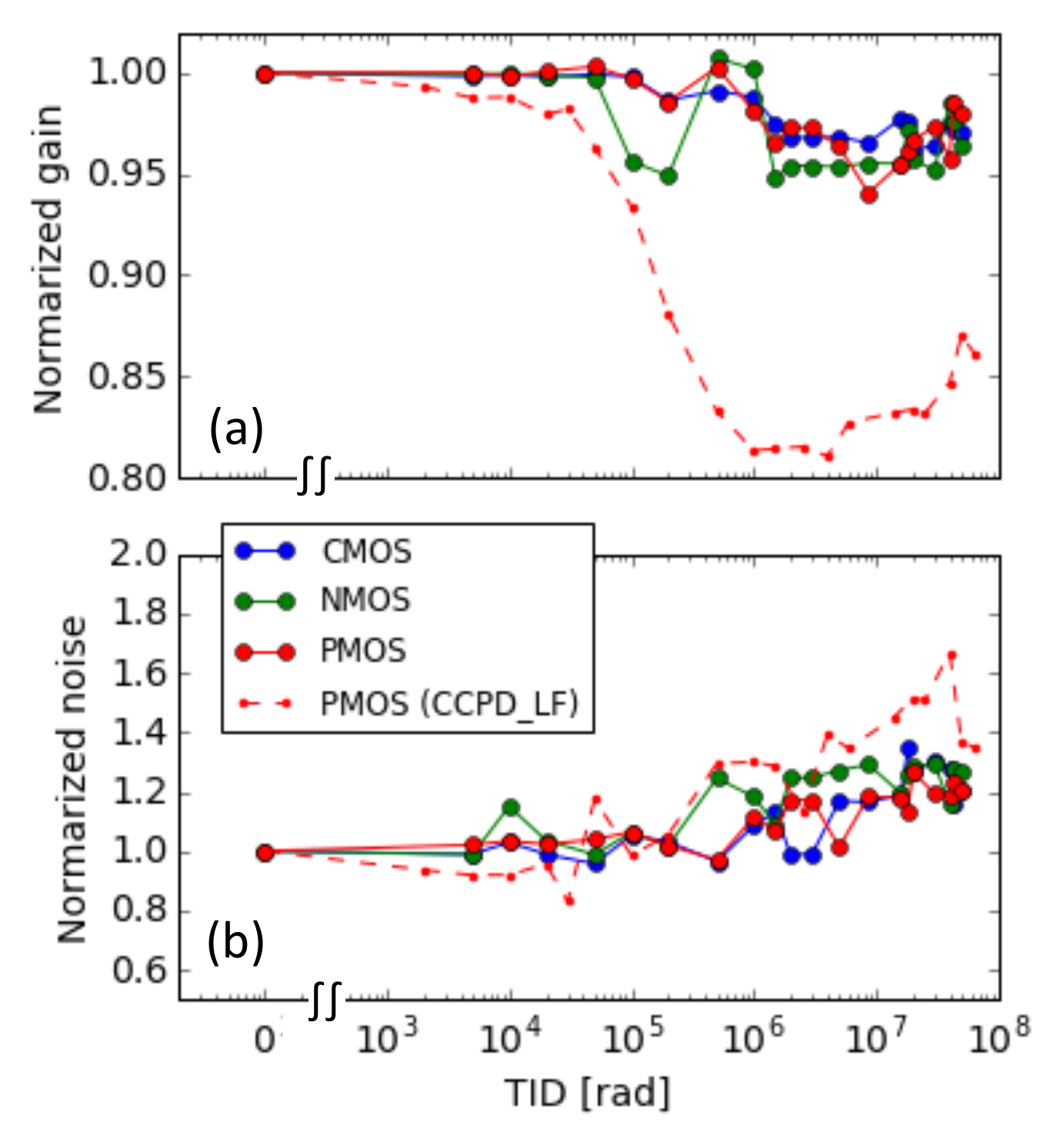}
    \caption[The gain and noise of LF-CPIX readout.]{
    Gain and noise of the LF-CPIX readout. The gain and noise were normalized to the values before irradiation.}
    \label{fig:gain}
\end{figure}

A test pulse injection method was used to obtain the gain and noise of the readout at each step. 
Figure \ref{fig:gain} (a) and (b) show that the gain decreases by about 5~\% and the noise increases by ~50~\%. 
Results of PMOS input transistor in the previous small prototype (CCPD\_LF) \cite{hirono:2015} is also shown in Figure \ref{fig:gain} for reference.
The circuit design of the LF-CPIX was modified from the CCPD\_LF to improve the TID radiation hardness because the gain degradation of the CCPD\_LF was as large as 20~\%.
Modifications included transistor sizes, layout, biasing circuitry at the CSA input, and peripheral circuitry outside the pixels. 
Although the most effective modification cannot be determined by this comparison, those improvements were effective.

The threshold dispersion of the discriminator was compared before and after 50~Mrad TID irradiation.
The un-tuned threshold dispersion increased from 370~e$^-$ to 420~e$^-$ due to the irradiation.
However, TDAC has the possibility to alleviate the increased threshold dispersion. 
An increase of only 20~e$^-$ is observed after TDAC tuning. 
This is negligible compared to the readout noise of 200~e$^-$.

\begin{figure}
    \centering   
    \includegraphics[width=0.8\linewidth,trim=0cm 15.0cm 5cm 0cm,clip]{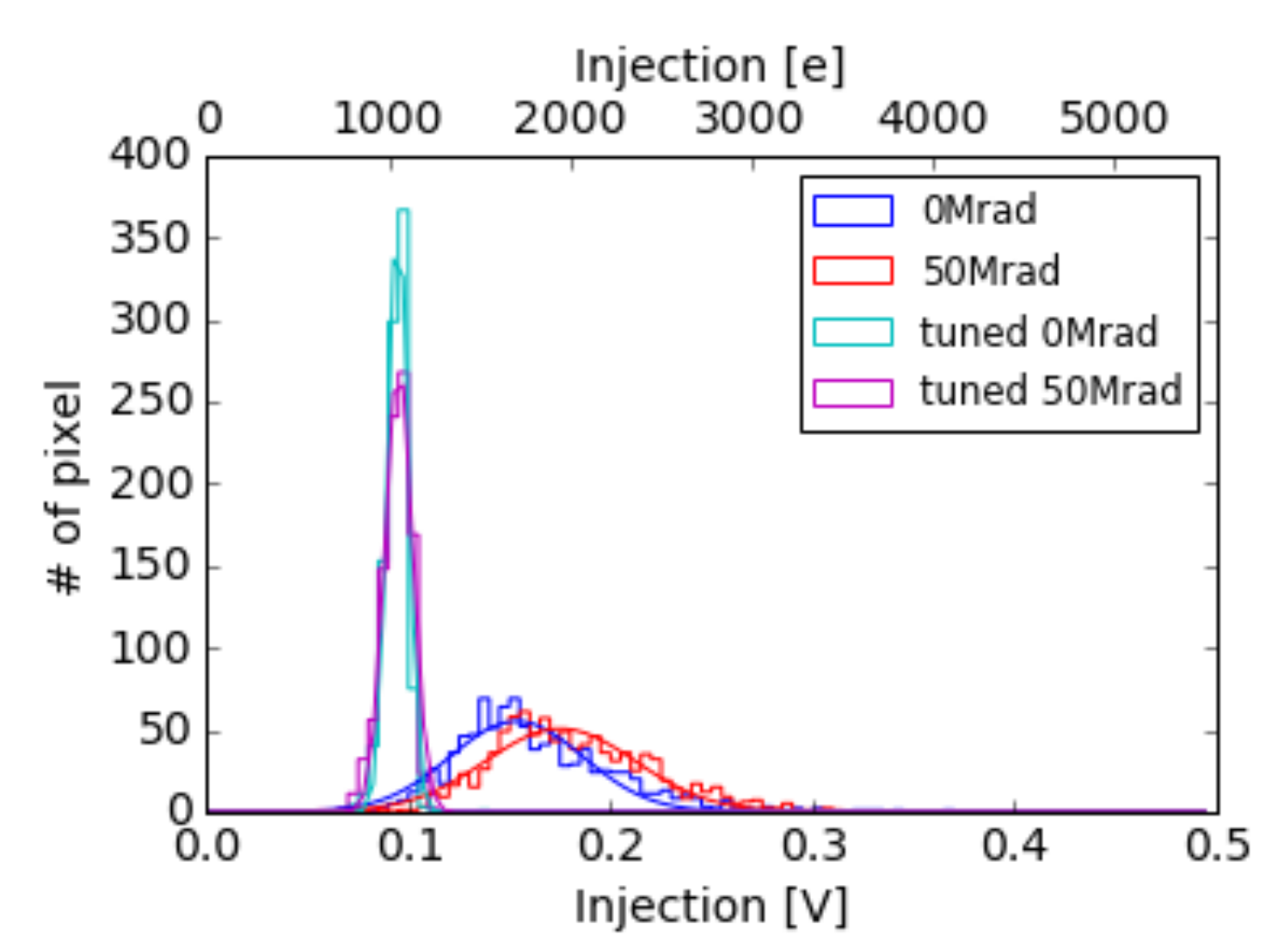}
    \caption[The threshold dispersion before and after irradiation with 50~Mrad TID]{
    The threshold dispersion before and after TID irrdiation of 50~Mrad (step lines).
    A Gaussian function is fitted to each dispersion (curved line).
    Threshold dispersion before irradiation without TDAC tuning, after irradiation without TDAC tuning, before irradiation with TDAC tuning, and after irradiation with TDAC tuning are 370, 418, 55, and 76~e$^-$, respectively.}
    \label{fig:th}
\end{figure}

\subsection{NIEL radiation hardness}
NIEL radiation damages the silicon crystal lattice and shortens the lifetime of charge carriers.
An LF-Monopix chip has been irradiated by neutrons with 10$^{15}$~n$_{\rm eq}$/cm$^2$ fluence in Jozef Stefan Institute reactor \cite{jsi}. 
Responses of the non-irradiated and irradiated chip to the 2.5~GeV electron beam have been compared using the fast readout, which fully functions even after irradiation. 
Measurements were performed at the External Beamline for Detector Tests (E3XD) in the Electron Stretcher Accelerator (ELSA) \cite{elsa:e3xd}. 

\begin{figure}
    \centering   
    \includegraphics[width=0.8\linewidth,trim=0cm 12cm 5cm 0cm,clip]{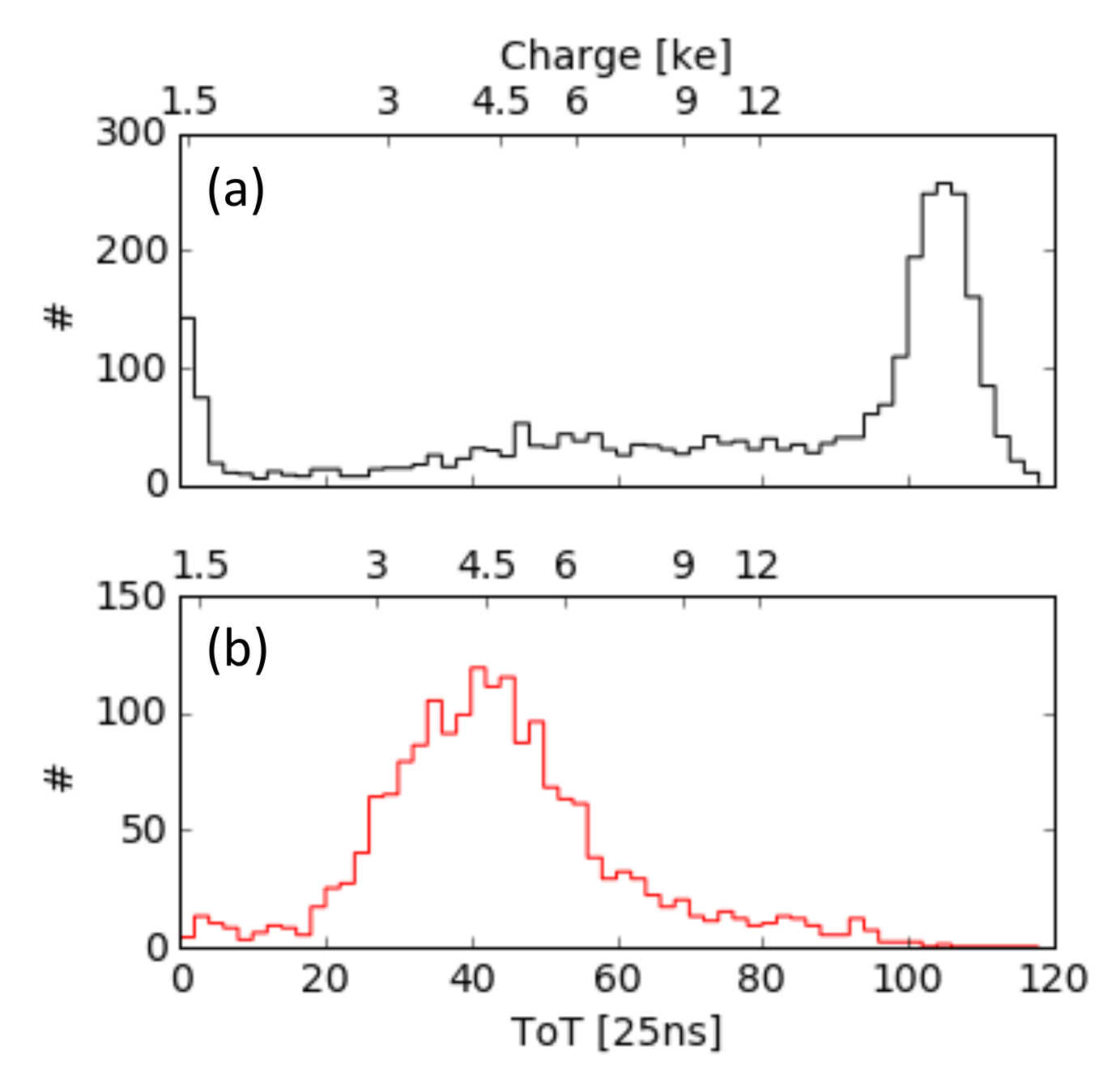}
    \caption[Charge spectrum of non-irradiated and irradiated LF-Monopix]{
    Charge spectrum of seed hits of the non-irradiated (a) and irradiated (b) LF-Monopix. 
    The fluence of the irradiated chip was 10$^{15}$~n$_{\rm eq}$/cm$^2$.
    The threshold was 1500~e$^-$. 
    The calibrated value in electrons is shown in the top axis of each plot.}
    \label{fig:charge}
\end{figure}

The charge spectrum is shown in Figure \ref{fig:charge}. 
The calibrated charge is shown in the upper axis up to 12~ke$^-$.
The ToT resolution for small signals is higher than that for large signals because distinguishing small hits from large hits is important in a tracker application.
Moreover, the gain calibration circuitry in LF-Monopix saturates around 15~ke$^-$ under the same conditions the charge spectrum measurements. 
The most probable value (MPV) of the non-irradiated chip is higher than the calibration range. 
The MPV of the irradiated chip is more than a factor of 5 smaller than that of the non-irradiated chips. 
However, it is still as high as 4.5~ke$^-$.

\begin{figure}
    \centering     
    \includegraphics[width=0.8\linewidth,trim=0cm 15cm 2.5cm 0cm,clip]{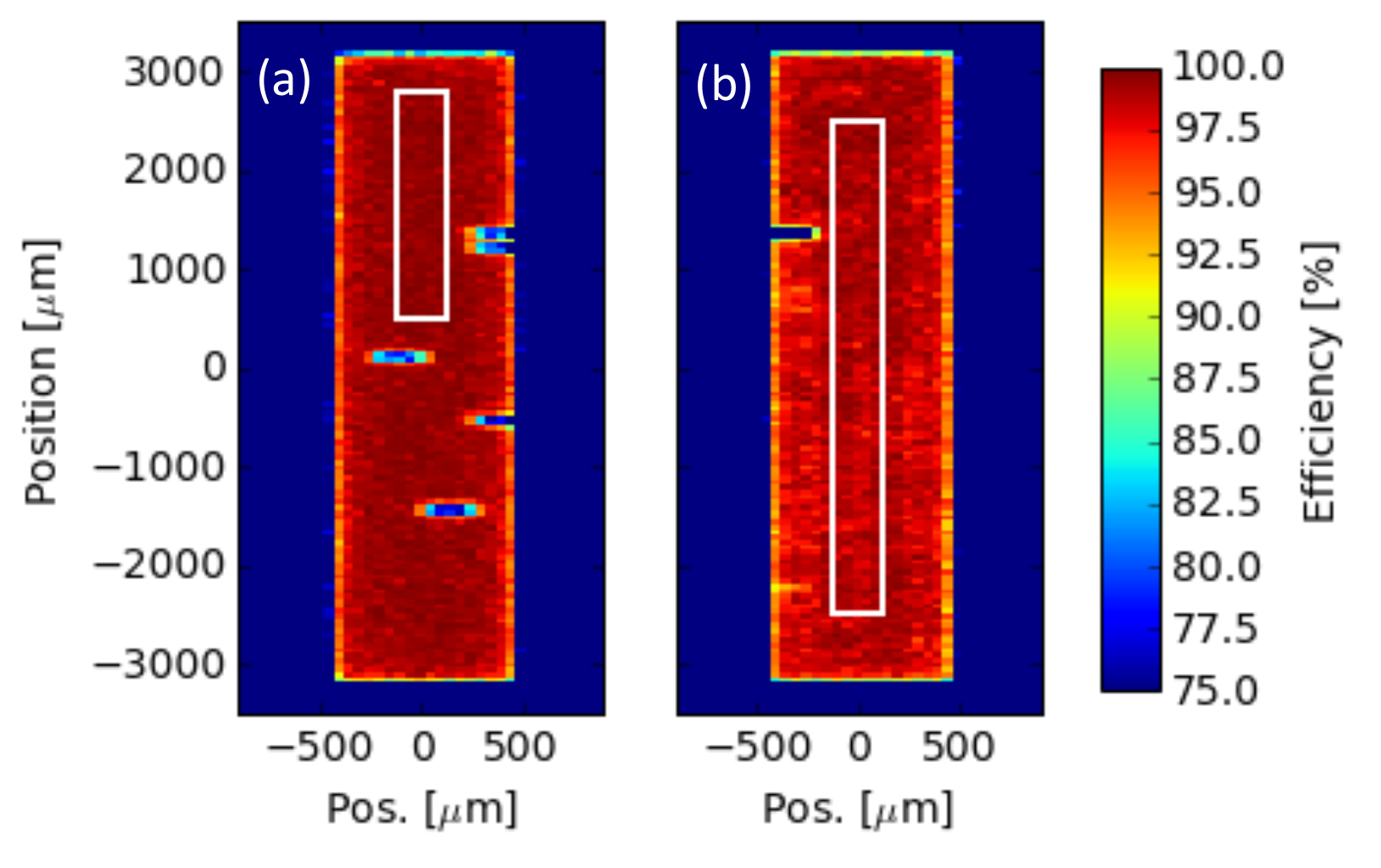}
    \caption[The hit-efficiency of non-irradiated and irradiated LF-Monopix]
    {The hit efficiency of the non-irradiated (a) and irradiated (b) LF-Monopix.
    The fluence of the irradiated chip was 10$^{15}$~n$_{\rm eq}$/cm$^2$.
    The white box indicates the region where the average hit efficiency was calculated}
    \label{fig:eff}
\end{figure}

The hit-efficiency was measured using an EUDET-type beam telescope \cite{eudet} with a newly developed fast readout system \cite{pascal:thesis}.
The beam telescope tracks the trajectory of the electron beam on LF-Monopix. 
The bias voltage on the non-irradiated and irradiated chip was set to 200~V and 130~V, receptively.
The threshold of the discriminator has been tuned to the median value of 1800~e$^-$ and 1600~e$^-$, respectively.
The noise occupancies at the thresholds were lower than 10$^{-7}$ per a clock period (25~ns), whereas noise occupancy is required to be less than 10$^{-6}$~/clk.
Both chips were cooled with dry ice, and the chip temperature was lower than -40~$^\circ$C. 
Impurities in the reconstructed track are estimated to be 0.5~\% and 0.3~\%, respectively.
These values limit the precision of the measured values.
The averaged hit efficiencies are 99.6~\% and 98.9~\%, respectively (Fig. \ref{fig:eff}).
A few noisy pixels are disabled, and the low efficiency regions correspond to those pixels.
The measured hit-efficiency is larger than the of ATLAS ITK pixel detector specifications ($\geq$~97~\%) \cite{atlas:pixel}. 

\section{Conclusion}
\label{sec:con}

The prototype chips demonstrate the suitability of the HR high large collection node DMPAS as ATLAS ITk pixel detector. 
The full depletion voltage of the 100~{\textmu}m thinned LF-CPIX was approximately around 15~V.
This value is far below the 220~V breakdown voltage.

The TID radiation hardness of the DMAPS has been tested up to a dose of 50~Mrad.
Gain degradation was suppressed compared to the previous prototype. 
The measurements show that the new CMOS-CSA is also radiation hardened. 
The NIEL radiation hardness  was investigated using the fully-monolithic large-scale prototype, which is equivalent to the current ATLAS pixel detector readout chip. 
The hit efficiency was sufficiently as high as 98.9~\% after NIEL irradiation with 10$^{15}$ n$_{\rm eq}$/cm$^2$ fluence, although a decrease in MPV was observed in the irradiated chip. 
Since thinning can yield a strong electric field, it is possible to recover the MPV by increasing the bias voltage and applying thinning to the DMAPS. A post-process thinning and back-side processing of DMAPS did not degrade the sensor properties.

\section*{Acknowledgment}
This work was supported in parts 
by the Deutsche Forschungsgemeinschaft DFG, grant number WE 976/4-1, 
by the German Ministry BMBF under grant number 05H15PDCA9, 
and by the H2020 project AIDA-2020, GA no. 654168, 
and by the H2020 project STREAM, GA no. 675587.

\section*{References}

\bibliography{mybibfile_short}

\end{document}